\documentclass[useAMS,usenatbib]{mn2e}

\usepackage{amsfonts,amsmath,amssymb,mathrsfs}
\usepackage{graphicx,epsf,epsfig}
\usepackage{bm}
\usepackage{longtable}
\usepackage[usenames,dvipsnames]{xcolor}
\usepackage{multirow}

\usepackage{graphicx}
\usepackage{hyperref}
\usepackage{times}
\usepackage{color}
\usepackage{breakurl}
\usepackage{color}
\usepackage{mathrsfs}

\def\be{\begin{equation}}
\def\ee{\end{equation}}
\def\beq{\begin{eqnarray}}
\def\eeq{\end{eqnarray}}
\def\bi{\begin{itemize}}
\def\ei{\end{itemize}}
\def\ben{\begin{enumerate}}
\def\een{\end{enumerate}}

\newcommand{\cf}{cf.,~}
\newcommand{\ie}{i.e.,~}
\newcommand{\eg}{e.g.,~}

\newcommand{\newtxt}[1]{{#1}}

\title[$I\,$--$\,$Love$\,$--$\,Q$ relations in magnetized neutron
stars]{On the universality of
  $\boldsymbol{I}\,$--$\,$Love$\,$--$\,\boldsymbol{Q}$ relations in
  magnetized\\ neutron stars}

\author[Haskell, Ciolfi, Pannarale, and Rezzolla]{B.~Haskell$^{1,2}$,
  R.~Ciolfi$^{1}$, F.~Pannarale$^1$, and L.~Rezzolla$^{1,3}$ \\
  \\
  $^1$ Max-Planck-Institut f\"{u}r Gravitationsphysik,
  Albert-Einstein-Institute, Am M\"{u}hlenberg 1, Potsdam, D-14776,
  Germany \\
  $^2$ School of Physics, The University of Melbourne, Parkville,
  Victoria 3010, Australia\\
  $^3$ Institut f\"ur Theoretische Physik, Max-von-Laue-Str. 1, 
  D-60438 Frankfurt, Germany
}

\begin{document}
\date{}
\maketitle
\begin{abstract}
  Recently, general relations among the quadrupole moment ($Q$), the
  moment of inertia ($I$), and the tidal deformability (Love number)
  of a neutron star were shown to exist. They are nearly independent
  of the nuclear matter equation of state and would be of great aid in
  extracting parameters from observed gravitational-waves and in
  testing general relativity. These relations, however, do not account
  for strong magnetic fields. We consider this problem by studying the
  effect of a strong magnetic field on slowly rotating relativistic
  neutron stars and show that, for simple magnetic field
  configurations that are purely poloidal or purely toroidal, the
  relation between $Q$ and $I$ is again nearly universal. However,
  different magnetic field geometries lead to different $I\,$--$\,Q$
  relations, and, in the case of a more realistic twisted-torus
  magnetic field configuration, the relation depends significantly on
  the equation of state, losing its
  universality. $I\,$--$\,$Love$\,$--$\,Q$ relations must thus be used
  with very great care, since universality is lost for stars with long
  spin periods, \ie $P \gtrsim 10\,$s, \emph{and} strong magnetic
  fields, \ie $B \gtrsim 10^{12}\,$G.
\end{abstract}


\section{Introduction}
\label{sec:intro}

Neutron stars (NSs) offer a unique opportunity to investigate the
state of matter at high densities, allowing us to probe aspects of the
strong interaction in conditions that cannot be reproduced with
terrestrial experiments. The equation of state (EOS) of nuclear matter
at such high densities is highly uncertain, and constraints can only
be inferred indirectly by studying its imprint on the exterior
properties of the star. For example, there have been recent efforts to
use observations of X-ray bursters to simultaneously constrain the
mass and the radius of the NS \citep{Steiner2010}, while radio
observations of pulsars, such as the double binary J0737-3039
\citep{Burgay03}, could be able to put constraints on the moment of
inertia. It is also likely that gravitational wave (GW) observations
of binary NS inspirals with Advanced LIGO/Virgo \citep{Harry2010},
KAGRA \citep{Somiya2011}, or the planned Einstein Telescope
\citep{Punturo2010b} will allow for further constraints on the spin,
quadrupolar deformation, and tidal Love number of the star.

Recent work has shown that, in slowly rotating and weakly magnetized NSs,
unique relations exist between the quadrupole moment, the moment of
inertia, and the tidal Love number. These relations are ``universal'', as
they are essentially independent of the EOS as first shown by \citet{Yagi2013a} and then confirmed by
 \citet{Maselli2013}, and could be used to break degeneracies between
parameters in GW signals. This would allow, for example, to determine NS
spins, conduct tests of general relativity \citep{Yagi2013b}, and
distinguish between NSs and strange stars \citep{Urbanec2013,Yagi2013a}.

In this paper, we consider the effect of the stellar magnetic field on
such universal relations. NSs are strongly magnetized stars, with
magnetic fields at the surface inferred to be of up to $10^{12}\,$G
for radio pulsars, and of up to $10^{15}\,$G for magnetars. It is well
known [see, \eg \citet{Chandrasekhar1953, Bocquet1995, Haskell2008,
  Ciolfi2010, Frieben2012}] that a magnetized NS cannot be spherical,
with deformations that can lead either to oblate or prolate shapes,
and may be even larger than those due to rotation.

First of all, we will show that, although the influence of the EOS is
weak for a given simple magnetic field configuration, different
geometries of the field lead to a different relation between the
quadrupolar deformation and the moment of inertia. Therefore, as soon
as deformations are dominated by magnetic fields the universality no
longer holds. Most NSs in binaries are spinning fast enough that this
is not the case. However, in slow enough systems, \ie when the NS spin
period is of the order of a few seconds, it is possible that the
quadrupolar deformation may be dominated by magnetic effects. It is
generally thought that the interior magnetic field of a NS may be much
stronger than the surface field \citep{Braithwaite2009, Corsi2011,
  Ozel2013}. Recent calculations of equilibrium models with magnetic
fields in a twisted-torus configuration support this view, showing
that the internal field can be up to two orders of magnitude stronger
than the external one, leading to very large deformations
\citep{Ciolfi2013}.  In slowly rotating stars these magnetic
deformations can easily dominate the quadrupole. Moreover, for
twisted-torus configurations the $I$-$Q$ relation depends on the EOS,
further invalidating the universality.

As a result, the ``universal'' relations found by \citet{Yagi2013a}
and extended by \citet{Maselli2013} are not applicable to highly
magnetized, slowly spinning NSs, for which they would lead to an
erroneous determination of the GW parameters. On the other hand,
measured deviations from the universal relations of \citet{Yagi2013a}
and \citet{Maselli2013} may potentially be used to constrain the
geometry of the NS internal magnetic field, which cannot be probed
with standard electromagnetic observations.

\section{Formalism}
\label{sec:formalism}

We calculate the relation between the quadrupolar deformation of the
star, $Q$, and its moment if inertia, $I$. We present our results in
terms of the dimensionless quantities $\bar{I} \equiv I/M^3$ and $\bar{Q}
\equiv Q/(M^3\chi^2)$, where $M$ is the mass of the star, and $\chi
\equiv J/M^2$, $J$ being the spin angular momentum of the star
\citep{Yagi2013a}\footnote{\newtxt{Different dimensionless normalizations
    are also possible, \eg in terms of
    $I/(MR^2)$~\citep{Lattimer01,Bejger02,Urbanec2013}.}}. Note that the
quadrupolar deformation is the result of a rotational part, $Q_{\rm r}$,
and a magnetic part, $Q_{\rm m}$, \ie $Q = Q_{\rm m} + Q_{\rm r}$, but
the normalization of $Q$ assumes that the star is always rotating, \ie
that $\chi \neq 0$.  Already for nonrotating models, however, $Q_{\rm m}
\neq 0$ in the presence of a magnetic field, and the natural quantity to
use to obtain a normalization would thus be the magnetic energy. For
simplicity, and to easily compare with previous results, we will continue
to define $\bar{Q}$ as $Q/(M^3\chi^2)$. Note that, for all practical
purposes, the slowly rotating models considered here have essentially the
same physical properties as the corresponding nonrotating ones.

In what follows we briefly discuss the general-relativistic
mathematical setups used for the calculation of $\bar{Q}$ and
$\bar{I}$, either within a perturbative approximation, or in a fully
nonlinear approach. To understand how a magnetic field can break the
universality of the $\bar{I}\,$--$\,\bar{Q}$ scaling relations,
however, it is instructive to first consider the much simpler
Newtonian case. It is sufficient to consider the Newtonian results for
a rotating polytropic star with polytropic index $n=1$ and polytropic
constant $\kappa=4.25\times 10^4$ cm$^5\,$g$^{-1}$ s$^{-2}$
\citep{Haskell2008}. At lowest order for a purely poloidal magnetic
field, the scaling relation between the normalized quadrupole and
moment of inertia is given by
\begin{equation}
\label{eq:Newt_pol}
\bar{Q}\approx 4.9\;{\bar{I}}^{1/2} + 10^{-3}\bar{I}\left(
  \frac{B_{p}}{10^{12}\,\mbox{G}}\right)^2
\left(\frac{P}{1\,\mbox{s}}\right)^2\,,
\end{equation}
where $B_{p}$ is the field at the pole, and $P$ the rotation
period. The first term in Eq.~(\ref{eq:Newt_pol}) is due to rotation
(\ie $\propto Q_{\rm r}$), while the second one is due to the
magnetization (\ie $\propto Q_{\rm m}$). This term was not analysed by
\citet{Yagi2013a} and \citet{Maselli2013}. Similarly, for a purely
toroidal field, the scaling relation is
\begin{equation}
\label{eq:Newt_tor}
\bar{Q}\approx 4.9\;{\bar{I}}^{1/2} - 3 \times 10^{-5}\bar{I}\left(
  \frac{\langle B \rangle}{10^{12}\,\mbox{G}}\right)^2
\left(\frac{P}{1\,\mbox{s}}\right)^2\,,
\end{equation}
where $\langle B \rangle$ is now the field averaged over the volume of
the star.

Given the expressions in (\ref{eq:Newt_pol}) and (\ref{eq:Newt_tor}), we
can make a number of remarks that will be valid also when considering the
results in a general-relativistic framework. First, in the case of purely
toroidal magnetic fields, the magnetic quadrupolar deformation is
negative, thus corresponding to a prolate shape. Second, with this
definition of $\bar{Q}$ the results depend on the product $B\times P$ and
will thus be, in general, ``non-universal'', as this product will vary
from system to system. We will thus investigate the effect of the EOS on
the $\bar{I}\,$--$\,\bar{Q}$ scaling relation at fixed period $P$, and
then study the effect of varying $P$. Finally, it is clear from the
coefficients in (\ref{eq:Newt_pol}) and (\ref{eq:Newt_tor}) that the
magnetic corrections are generally smaller than those associated with the
rotation, and that magnetic effects will only dominate for long rotation
periods and strong magnetic fields. Hereafter we will focus on the
$\bar{I}\,$--$\,\bar{Q}$ relation, since the corrections on the Love
number would be of higher order and no formulation of the Love number for
magnetized and rotating objects has been derived yet. \newtxt{It is
  clear, however, that a loss of universality in the
  $\bar{I}\,$--$\,\bar{Q}$ relation implies a loss of universality also
  in terms of the Love number.}

Let us now consider stellar equilibria in full general relativity, but
with magnetic fields that are either purely poloidal or purely
toroidal. Configurations of this type have been extensively studied in
the past \citep{Bocquet1995, Cardall2001, Kiuchi2008,
  Frieben2012}. Equilibrium models even with ultra-strong magnetic fields
can be readily computed via the publicly available \texttt{LORENE}
library\footnote{\url{http://www.lorene.obspm.fr}}, and we refer to
\citet{Bocquet1995} \newtxt{(\texttt{Magstar} code)} and
\citet{Frieben2012} for details on the numerical implementation in the
case of purely poloidal and purely toroidal configurations,
respectively. Although fully nonlinear and simpler to compute, these
purely poloidal or purely toroidal configurations are known to be
dynamically unstable on an Alfv\'en timescale
\citep{Markey1973}. Furthermore, the occurrence of this instability has
been verified in a number of recent nonlinear general-relativistic
simulations~\citep{Lasky2011, Ciolfi2011, Kiuchi2011, Ciolfi2012,
  Lasky2012}.

Let us thus consider a more realistic field topology, the so-called
``twisted-torus''.  In these configurations, the magnetic field has
both poloidal and toroidal components, with the toroidal being
possibly much stronger than the poloidal surface field. No
general-relativistic solution has yet been found for this
configuration in a fully nonlinear setup. Nevertheless, twisted-torus
configurations have been explored extensively in recent years, either
in Newtonian nonlinear equilibria \citep{Tomimura2005,
  Yoshida2006,Lander:2009}, or within general-relativistic
perturbative approaches \citep{Ciolfi2009,
  Ciolfi2010,Ciolfi2013}. Following the latter approach, we consider
the magnetic field as a perturbation on a background equilibrium
solution of a nonrotating star with an EOS $p=p(e)$, where $p$ is the
pressure and $e$ the energy density. Note that using nonrotating
background models is a good approximation for rotation periods $P
\gtrsim 10\,$s if the surface magnetic fields are $\gtrsim
10^{12}\,$G. \newtxt{More precisely, we find that $Q=Q_{\rm r}$ for 
fully relativistic rotating stars with $P \sim 10\,$s and $B=0$ 
is comparable to $Q=Q_{\rm m}$ for a twisted-torus 
configuration with $B_{p} \sim10^{12}\,$G}; these can be taken as 
the critical periods and magnetic fields such that $Q_{\rm r} \sim 
Q_{\rm m}$ for our twisted-torus configurations.

\pagebreak

The azimuthal component of the vector potential
$A_\phi=\psi(r,\theta)$ must satisfy the Grad-Shafranov equation
\begin{eqnarray}
  \frac{e^{-\lambda}}{4\pi}\left[\partial_r^2\psi + 
    \frac{\partial_r\nu-\partial_r\lambda}
    {2}\partial_r\psi\right]\!+\!\frac{\left(\partial_\theta^2\psi - 
      \cot\theta\partial_\theta\psi\right)}{4\pi r^2}\!=
  \nonumber\\
  =-\frac{e^{-\nu}}{4\pi}\beta\frac{d\beta}{d\psi}-F(e + p)r^2\sin^2\theta\,,
\end{eqnarray}
where the metric functions $\nu(r)$ and $\lambda(r)$ are determined
from the background solution, while $\beta(\psi)$ and $F(\psi)$ are
two arbitrary functions that determine the field geometry ($r$ and
$\theta$ are spherical coordinates). Once a solution for $\psi$ is
found by assuming regularity at the centre of the star and matching to
the external vacuum solution (taken to be dipolar for simplicity, \ie
$\psi(r,\theta)=-a_1(r)\sin^2\theta$), the magnetic field components
are obtained by taking the curl of the vector potential $A^i$. To
study twisted-torus configurations in which the toroidal magnetic
field can be comparable to or stronger than the poloidal surface
field, we consider the form suggested by \citet{Ciolfi2013} for the
trial functions, \ie
\begin{align}
  \beta(\psi)&=\zeta_0\psi(|\psi/\bar{\psi}|-1)\, \Theta(|\psi/\bar{\psi}|-1)
  \,,\label{ciolfi1}\\
  F(\psi)&=c_0\left[(1-|\psi/\bar{\psi}|)^4 \, \Theta(1-|\psi/\bar{\psi}|)-
    \bar{k})\right]\,,\label{ciolfi2}
\end{align}
%
%
\begin{figure}
  \centerline{\includegraphics[width=8cm]{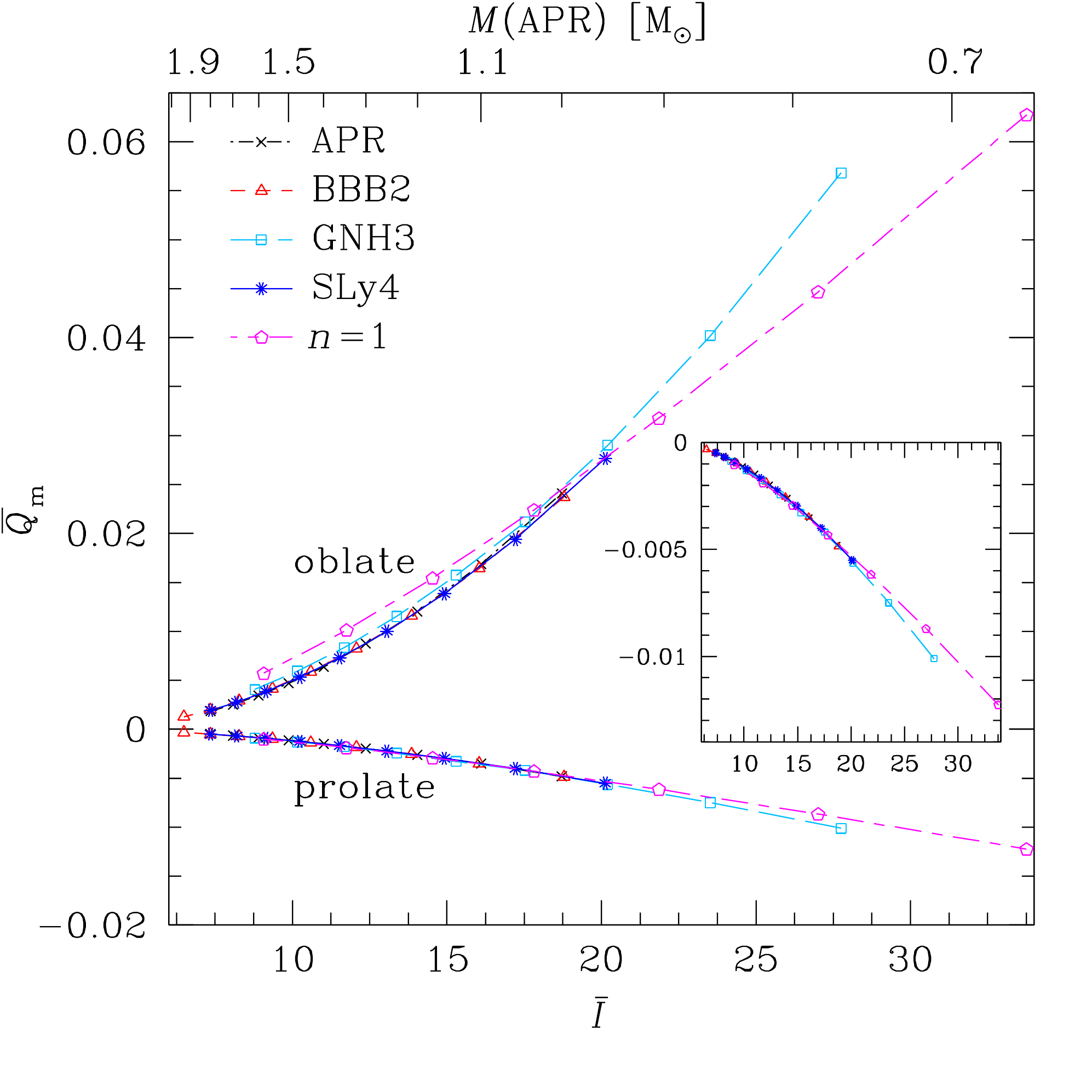}}
  \caption{Dimensionless magnetically induced quadrupole deformation,
    $\bar{Q}_{\rm m}$, as a function of the dimensionless moment of
    inertia, $\bar{I}$, for NSs with a purely poloidal magnetic field
    $B_{p}=10^{12}\,$G (oblate configurations, $\bar{Q}_{\rm m}>0$), and
    for NSs with a purely toroidal magnetic field $\langle B
    \rangle=10^{12}\,$G (prolate configurations, $\bar{Q}_{\rm m}<0$).
    The stellar mass for the APR EOS is shown in the upper $x$-axis as a
    reference; the models are nonrotating but $P=2\pi\,$s was used for
    $\bar{Q}$.}
  \label{tscales}
\end{figure}
%
where $\zeta_0$, $c_0$, and $\bar{k}$ are constants, $\bar{\psi}$ is
the value of $\psi$ on the last closed-field line (tangent to the
surface), and $\Theta(x)$ is the Heaviside step function. Once the
magnetic field configuration is determined, the new equilibrium
configuration is found by perturbing the continuity equation
$\nabla_{\mu}(nu^\mu) = 0$ and the relativistic equations of
hydrostatic equilibrium in the presence of electromagnetic fields

\begin{equation}
  (e+p)u^{\nu} \nabla_{\nu} u_{\mu}+ \partial_{\mu} p + u_\mu
  u^\nu \partial_\nu p = \frac{F_{\mu\nu} \nabla_{\alpha} F^{\nu\alpha}}{4\pi}\,,
\end{equation}
where $F_{\mu\nu} = \partial_\nu A_\mu-\partial_\mu A_\nu$ is the
Maxwell tensor, such that $F_{\mu\nu} u^\nu=0$, and $\nabla$ denotes
covariant derivatives with respect to the background metric. $I$ and
$Q$ are then determined by matching the interior solution to the
exterior metric of a slowly rotating star.

\begin{figure}
\centerline{\includegraphics[width=8cm]{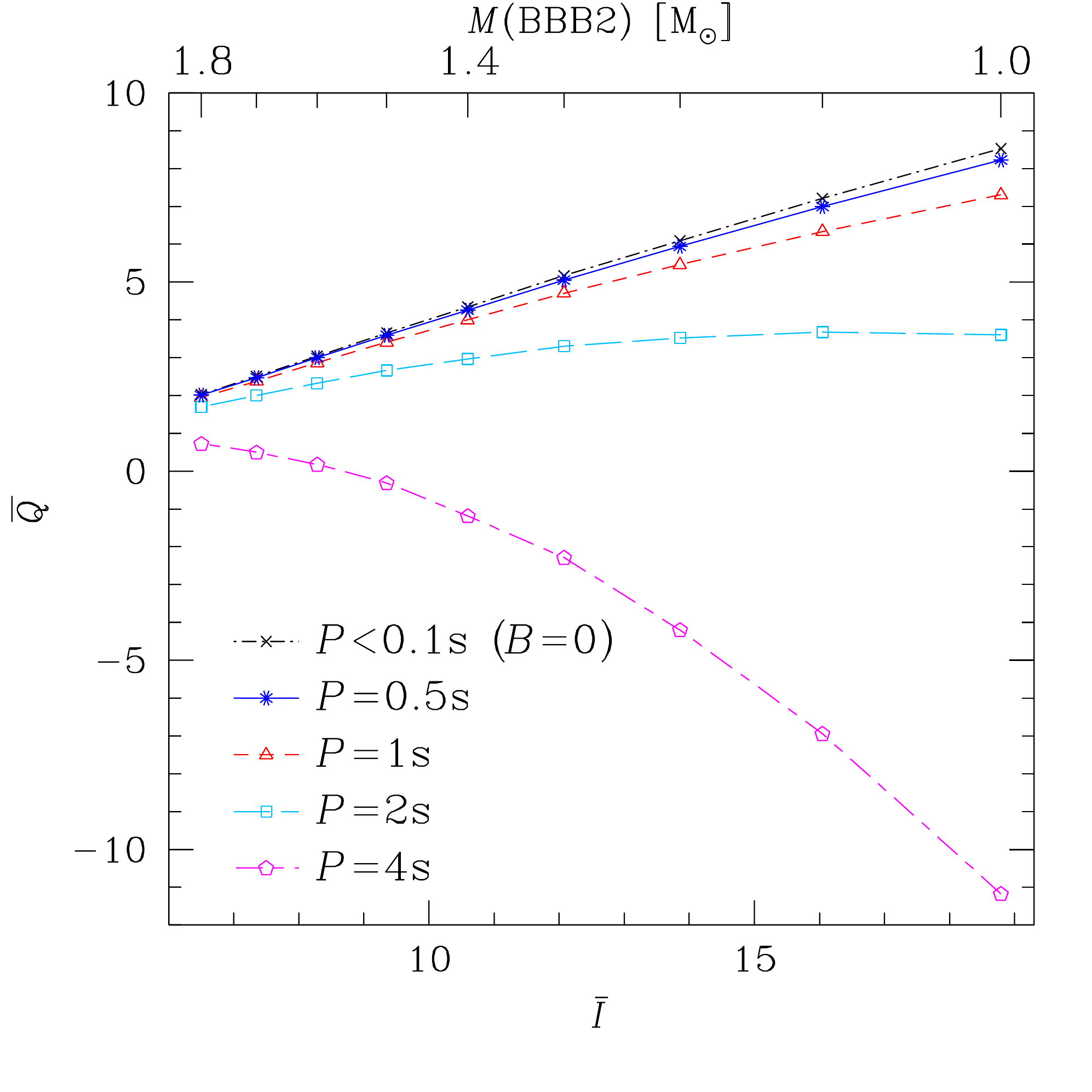}}
\caption{Relation between $\bar{Q}$ and $\bar{I}$ for a purely
  toroidal field with an internal field strength of $\langle B
  \rangle=10^{14}\,$G, for various rotation rates and the BBB2
  EOS. The stellar mass is shown in the upper $x$-axis. Note that
  the slope of the curve changes, increasing with the rotation rate as
  the star goes from being prolate to being oblate.}\label{rotation}
\end{figure}

\section{Magnetic deformations}
\label{sec:magdef}
\subsection{Purely poloidal and purely toroidal configurations}

\begin{figure*}
\centerline{
\includegraphics[width=8cm]{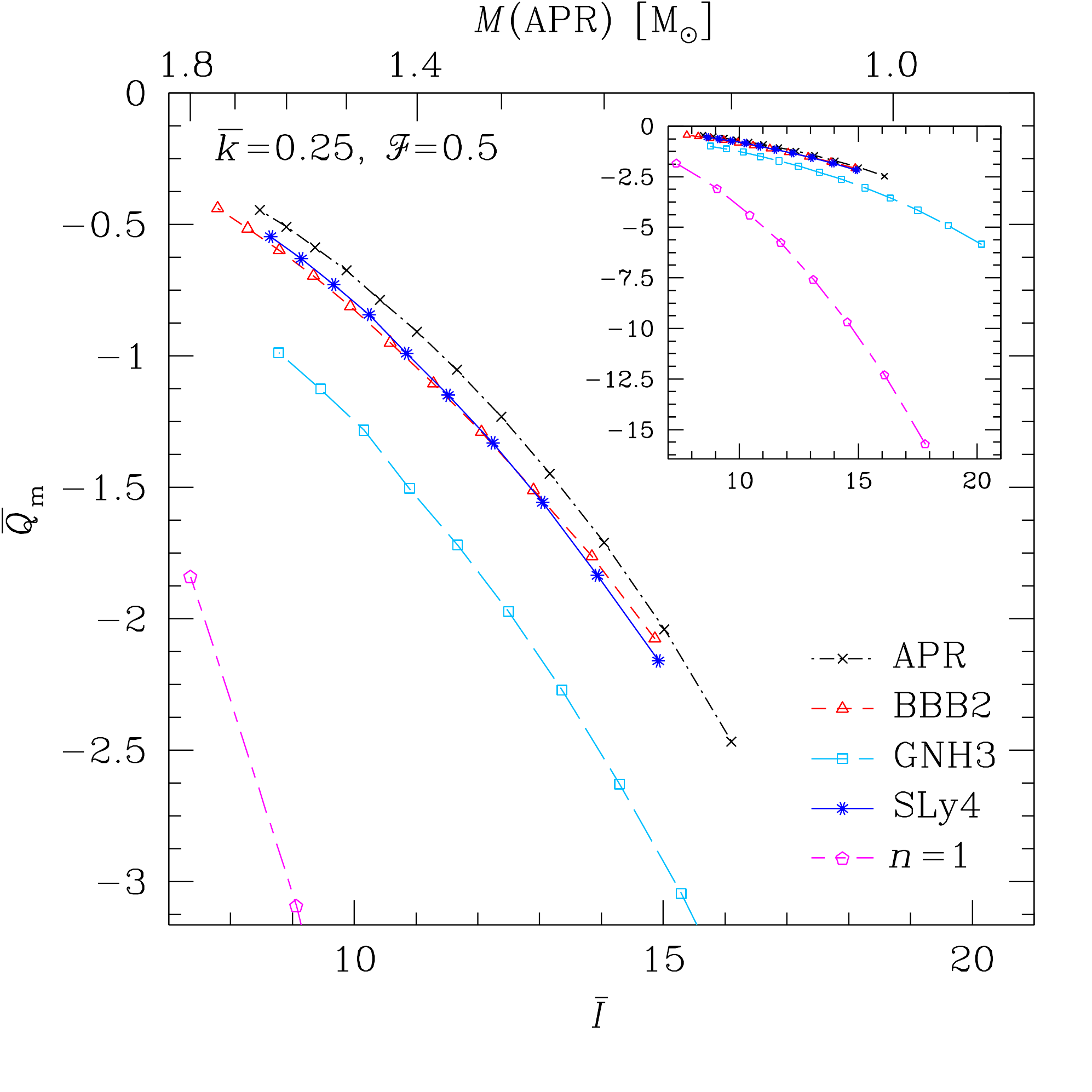}
\hskip 1.0cm
\includegraphics[width=8cm]{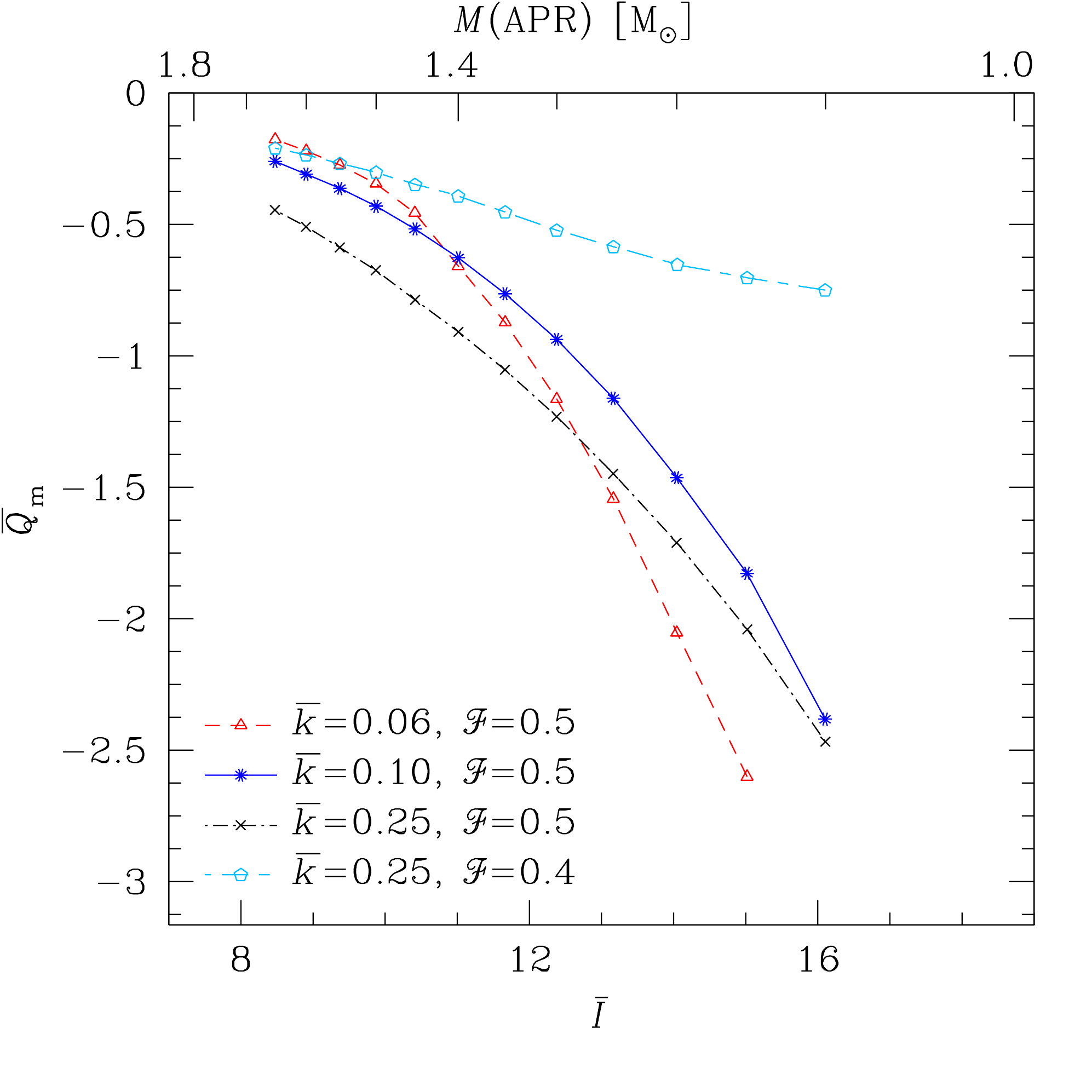}
}
\caption{\textit{Left panel:} Relation between $\bar{Q}_{\rm m}$ and
  $\bar{I}$ for twisted-torus configurations with $\mathscr{F}=0.5$,
  $B_{p}=5\times 10^{12}\,$G, $\bar{k} = 0.25$, and $P=10\,$s. Note that the
  star is prolate because of the strong internal toroidal field, and that
  the $\bar{I}$--$\bar{Q}$ relation is no longer recovered. \textit{Right
    panel:} The same as in the left panel for the APR EOS, with different
  sequences referring to different values of $\bar{k}$ or different
  fractions of the toroidal energy. In both panels the stellar mass for
  the APR EOS is shown in the upper $x$-axis as a reference.}
\label{ttorus}
\end{figure*}

We start our analysis by examining the case of a NS endowed with a
purely poloidal magnetic field. As mentioned above, such a
configuration is known to be unstable, but represents an adequate
first step towards examining the more realistic twisted-torus
configurations. We impose the field configuration using the
prescription by \citet{Bocquet1995} and calculate the quadrupolar
distortion of the star using the \texttt{LORENE} library. In
Fig.~\ref{tscales} we show the magnetically induced $\bar{Q}_{\rm m}$
as a function of $\bar{I}$, for $B_{p}=10^{12}\,$G. The data refers to
nonrotating models, but a period of $P=2\pi\,$s was used in
normalizing $Q$. The results reported refer to a polytropic EOS with
$n=1$ and $\kappa=14.6\times 10^4$ cm$^5\,$g$^{-1}$ s$^2$, and to four
realistic EOSs of cold nuclear matter, namely, the APR
EOS~\citep{Akmal1998a}, the BBB2 EOS~\citep{Baldo1997}, the GNH3
EOS~\citep{Glendenning1985}, and the SLy4 EOS~\citep{Douchin01}. In
this case, all the deformations lead to $\bar{Q}_{\rm m} > 0$, \ie to
an oblate shape, and it is easy to see that once the magnetic field
configuration is fixed, the relation between $\bar{I}$ and $\bar{Q}$
is fairly ``universal'' and depends only weakly on the EOS (the larger
differences for the polytrope are mostly due to its inaccurate
treatment of densities close to the crust).

A similar result holds in Fig.~\ref{tscales} when we consider purely
toroidal magnetic fields calculated using the prescription of
\citet{Frieben2012} and fixing the average magnetic field to $\langle
B \rangle=10^{12}\,$G. In this case, all deformations lead to a
prolate shape, \ie $\bar{Q}_{\rm m} < 0$, but, as for purely poloidal
fields, the relation appears to be EOS independent (the inset provides
a magnified view for prolate models). It is important to stress,
however, that such universality only holds once the magnetic field
configuration and strength are fixed. In general, different magnetic
field strengths lead to curves with different slopes, depending on the
product $\langle B \rangle \times P$.

The dependence on $\langle B \rangle \times P$ is evident in
Fig.~\ref{rotation}, where the total quadrupole $\bar{Q}$, \ie
including rotational deformations, is shown as a function of $\bar{I}$
for various rotation rates and an average magnetic field $\langle B
\rangle=10^{14}\,$G. Given the near universality of the results
discussed previously, we restricted the analysis to the BBB2 EOS, but
similar results hold for other EOSs. Note that as the rotation period
$P$ decreases from $4\,$s to $ < 0.1\,$s, from bottom to top, the
slope of the scaling relation increases as the star goes from being
prolate to being oblate. Different stars rotate at different rates and
have different magnetic field strengths. The product $\langle B
\rangle \times P$ will thus change from system to system, leading to
different $\bar{I}\,$--$\,\bar{Q}$ relations.  For very short periods,
however, the curves no longer show any influence of the magnetic
deformation and agree with the $\bar{I}\,$--$\,\bar{Q}$ relation for
an unmagnetized star, as the quadrupolar distortion is now dominated
by rotational effects. As a result, for periods below $\approx 0.1\,$s
the magnetic contribution to $\bar{Q}$ becomes negligible.

\subsection{Twisted-torus configuration}

We can now advance our analysis by considering more realistic
twisted-torus configurations. We use
Eqs.~(\ref{ciolfi1})--(\ref{ciolfi2}) to obtain twisted-torus
configurations with internal toroidal-to-total magnetic field energy
ratio $\mathscr{F} \equiv E_{\mathrm{tor}} /
E^{\,\mathrm{int}}_{\mathrm{m}} = 50\%$, a surface (polar) magnetic
field strength of $5\times 10^{12}\,$G, and $\bar{k}=0.25$ [\cf
Eq.~(\ref{ciolfi2})]; while the surface field is fixed, the interior
one changes from configuration to configuration, but is always
$\langle B \rangle \gtrsim 10^{13}\,$G.

The results for the $\bar{I}\,$--$\,\bar{Q}$ relation are summarized in
the left panel of Fig.~\ref{ttorus}, using different EOSs and a
$P=10\,$s normalization, (\cf Fig.~\ref{tscales}). In this case the
change in EOS has a considerable impact on the current distributions, and
this leads to significant differences in both $\bar{Q}$ and
$\bar{I}$. Although the realistic EOSs show a behaviour that does not
produce large variations (especially when compared to the $n=1$
polytrope, shown fully in the inset) the near universality of the
$\bar{I}\,$--$\,\bar{Q}$ relation found in Fig.~\ref{tscales} for purely
toroidal or poloidal configurations is not recovered. \newtxt{Note that
  although small, $\bar{Q}_{\rm m}$ is effectively comparable with
  $\bar{Q}_{\rm r}$ at these rotation rates and magnetic-field
  strengths.}

In addition, the results depend sensitively on the changes of the
overall poloidal-to-toroidal field ratio and on the prescription for
the currents inside the star. These changes lead to significantly
different geometries and relations between $\bar{Q}$ and $\bar{I}$, as
shown in the right panel of Fig.~\ref{ttorus}. In this panel, the
different sequences refer to different values of the parameter
$\bar{k}$, which in turn lead to different internal field
strengths. \newtxt{Note that once rescaled to the same polar magnetic 
field strength and rotation period, the twisted-torus deformations 
(\cf Fig.~\ref{ttorus}) are larger than those obtained with a purely 
poloidal field (\cf Fig.~\ref{tscales}) and the shape of the star is prolate, 
rather than oblate.} 
This is due to the strong internal toroidal component of the
magnetic field, which dominates the distortion for slowly rotating
NSs. Finally, as comparison we also show in the right panel of
Fig.~\ref{ttorus} a sequence still having a surface magnetic field
strength of $5\times 10^{12}\,$G and $\bar{k}=0.25$, but where $40\%$ 
of the magnetic energy is in the (internal) toroidal magnetic field (\ie
$\mathscr{F}=0.4$). Also in this case the new curve has a different
slope, as the poloidal contribution counters the toroidal one, leading
to less prolate configurations.

Note that we have not included the effect of superconductivity in our
analysis. The protons in the outer core are, however, expected to form
a type II superconductor, and this can substantially alter the
dynamics of the system, as the magnetic field will be expelled from
the bulk of the fluid and confined to flux tubes. Simple estimates
suggest that the quadrupole for a superconducting star, $Q_s$, is
simply related to that of a ``normal'' star, $Q_n$, by $Q_s\approx Q_n
H_1/B$, where $H_1$ is the lower critical field for superconductivity,
expected to be around $H_1\approx 10^{14}\,$G
\citep{Jones1975b,Easson1977}. While this simple scaling is
approximately true for purely poloidal and toroidal magnetic fields,
leading to even larger deformations than those discussed so far, the
situation for general mixed poloidal/toroidal fields is generally more
involved \citep{Lander2013,Lander2013b} and will be the focus of
future work.

\section{Conclusions}
\label{sec:conclusions}

We have shown that once a purely poloidal or a purely toroidal
magnetic field configuration is fixed, the relation between the
normalized magnetic quadrupole $\bar{Q}$ and the normalized moment of
inertia $\bar{I}$ is nearly ``universal'' and depends only weakly on
the EOS, in agreement with similar conclusions reached by
\citet{Yagi2013a} and \citet{Maselli2013} in the absence of magnetic
fields. However, if a more realistic twisted-torus configuration is
considered, in which poloidal and toroidal components coexist, the
field configuration itself depends on the EOS and could lead to
significant differences also in the $\bar{I}\,$--$\,\bar{Q}$ relation
for different EOSs. In general, different magnetic field geometries
and/or strengths could lead to a different relation, even for the same
EOS. Furthermore, already the Newtonian estimates
(\ref{eq:Newt_pol})--(\ref{eq:Newt_tor}) show that the value of
$\bar{Q}$ depends also on the ratio between magnetic and rotational
energies, thus differing from star to star.

Naturally, a departure from universality in the
$\bar{I}\,$--$\,\bar{Q}$, and hence $I\,$--$\,$Love$\,$--$\,Q$,
relation will have strong implications for GW detection. It will no
longer be possible to use the universal relations derived by
\citet{Yagi2013a,Yagi2013b} and \citet{Maselli2013} to reduce the
parameter space to search, and any parameter inferred from them will
not be reliable unless it is known that the stars have weak magnetic
fields, \ie $B \lesssim 10^{12}\,$G, and are rotating at periods $P
\lesssim 10\,$s. Above these periods \emph{and} magnetic field
strengths the universality is lost when considering twisted-torus
configurations.

Luckily, for most binary NS systems of interest for GW detection, it
should be possible to use the $I\,$--$\,$Love$\,$--$\,Q$ relations,
but not for all. For systems with more slowly rotating components this
will require extreme care. As an example, let us consider the
so-called ``double'' pulsar PSR J0737-3039. This is a binary NS system
in which both NSs are seen as radio pulsars \citep{Lyne04}.  Pulsar A
has a spin period of $P=22.7\,$ms and an estimated field strength of
$6.3\times 10^9\,$G. Pulsar B is much slower and has a spin period of
$P=2.77\,$s, with an estimated field strength of $1.2\times
10^{12}\,$G. The time to merger is estimated to be around $85\,$Myrs,
at which point the spin period of pulsar B will have slowed down to
$P\approx 3.9\,$s (assuming standard electromagnetic spin down and no
field decay). The results of \citet{Ciolfi2013} suggest that a
realistic NS could plausibly harbour a strong internal magnetic field
(up to $2$ orders of magnitude stronger than the surface field),
potentially leading to a situation very similar to the one illustrated
in Fig.~\ref{rotation}, where for $P\approx 4\,$s, the value of the
quadrupole deviates significantly from that of an unmagnetized
rotating star, and the NS could even be prolate. In this particular
system the average quadrupole is dominated by the rotational
contribution of pulsar A, but great care must be used in systems
containing such slowly rotating stars.

Such deviations from a universal relation would also hinder any test
of general relativity, as they would introduce many more parameters in
the analysis, and deviations from the expected trend could be
prescribed to an unobserved strong interior magnetic field
component. On the other hand, independent measurements of the
different quantities (such as $\bar{Q}$ and $\bar{I}$) could lead to
the identification of a strong internal magnetic field. We note that
to leading order in the post-Newtonian analysis, the magnetic field
would not affect the tidal deformability (i.e. the Love number), but
it would impact on higher order corrections. The presence of a
magnetic field, in fact, selects a preferred direction in space. As a
result, the deformability of the star under an external tidal field is
affected in a way which depends on the magnetic field strength, on its
topology, and, ultimately, on the EOS. These ``orientation
corrections'', which could be misinterpreted as a highly multipolar
magnetic field, will affect the emitted GWs in a way which has so far
not been quantified. Neglecting these corrections could lead to an
erroneous determination of the system parameters. It is thus essential
that the $I\,$--$\,$Love$\,$--$\,Q$ relations are used with great
care, ensuring that the spin of the stars is sufficiently rapid and
the magnetic field sufficiently weak, so that the relations can be
applied with confidence.

\medskip
\noindent
We thank J.~Frieben for providing the data in Figs.~\ref{tscales} and
\ref{rotation} and for his support. We are grateful to V.~Ferrari,
L.~Gualtieri, A.~Maselli, K.~Yagi, and N.~Yunes for useful comments. RC
is supported in part by the Humboldt Foundation and BH by the ARC via a
DECRA fellowship. Support comes also from the DFG grant SFB/Transregio~7
and from ``CompStar'', an ESF Research Networking Programme.

\vspace{-0.5cm}
\bibliographystyle{mn2e}
\bibliography{aeireferences}

\end{document}